%% file: paper.tex
\documentclass[lettersize,journal]{IEEEtran}
\usepackage{amsmath,amsfonts}
\usepackage{algorithmic}
\usepackage{algorithm}
\usepackage{array}
\usepackage[caption=false,font=normalsize,labelfont=sf,textfont=sf]{subfig}
\usepackage{textcomp}
\usepackage{stfloats}
\usepackage{url}
\usepackage{verbatim}
\usepackage{graphicx}
\usepackage{cite}
\usepackage{subfiles}
\usepackage{graphicx}
\usepackage{amsmath}
\usepackage{amssymb}
\usepackage{multirow}
\usepackage{makecell}
\usepackage{color}
\usepackage{bm}
\usepackage[pagebackref,breaklinks,colorlinks]{hyperref}
\usepackage{marvosym}
\usepackage{xcolor}
\usepackage{booktabs}
\usepackage{colortbl}
\usepackage{threeparttable}

\usepackage[capitalize]{cleveref}
\crefname{section}{Sec.}{Secs.}
\Crefname{section}{Section}{Sections}
\Crefname{table}{Table}{Tables}
\crefname{table}{Tab.}{Tabs.}

\input{./includes/newcommands.tex}

\newcommand{\TODO}[1]{\colorbox{yellow}{TODO}}

\newcommand{\NEWHALFLINE}{\vspace{0.5em}\noindent}

\newcommand{\citep}[1]{\cite{#1}}
\newcommand{\citet}[1]{\cite{#1}}

\begin{document}

\title{
    From State of THE ART to Human Prediction Algorithms
}

\author{
    Conghao Wong,
    Ziqian Zou
    \thanks{
        The authors are with Huazhong University of Science and Technology, China
        (Email: conghaowong@icloud.com,
        ziqianzoulive@icloud.com).
    }
}

\markboth{Journal of \LaTeX\ Class Files,~Vol.~14, No.~8, August~2021}%
{Shell \MakeLowercase{\textit{et al.}}: A Sample Article Using IEEEtran.cls for IEEE Journals}


\maketitle

\subfile{./contents/0_abstract.tex}


\subfile{./contents/1_sotas/c1s1.tex}

\subfile{./contents/2_art/c2s1.tex}

\subfile{./contents/2_art/c2s2.tex}

\subfile{./contents/2_art/c2s3.tex}

\subfile{./contents/2_art/c2s4.tex}

\subfile{./contents/3_unpredictable/c3s1.tex}

\bibliographystyle{IEEEtran}
\bibliography{ref.bib}



\vfill

\end{document}

%% file: includes/newcommands.tex
\newcommand{\SECTION}[1]{\cref{#1}}

\newcommand{\ETAL}{{\emph{et al.}}}

\newcommand{\IE}{{\emph{i.e.}}}

\newcommand{\SCMODEL}{SocialCircle}

\newcommand{\REMODEL}{\emph{Resonance}}

\newcommand{\REVMODEL}{\emph{Reverberation}}

%% file: contents/0_abstract.tex
\begin{abstract}

Human curiosity and longing for the future have driven human progress from ancient times to the present.
With the advancement of AI technology, human prediction algorithms have also drawn increasing attention from researchers.
However, as AI technology continues to advance, we observe a counterintuitive phenomenon.
Take human trajectory prediction algorithms as an example.
Although models' quantitative prediction performance has improved, yielding one state-of-the-art model after another, the consideration of human behaviors, future thoughts, and social interactions themselves presents a gradually diminishing trend in the construction of these prediction algorithms, buried beneath numbers and metrics.
In this manuscript, taking interactive art and human trajectory prediction algorithms as examples, we attempt to adopt a different analytical perspective to analyze the structural and developmental commonalities between these two fields.
By examining specific interactive artworks and prediction algorithms, we aim to explore how human intuitions and art are associated with AI algorithms through three aspects, including perception, interaction, and timing, thereby providing new insights for the understanding and further research in this domain.
We also observe an interesting phenomenon that certain art-inspired prediction algorithms can actually deduce intriguing patterns of human behaviors from the data, offering new potential avenues for understanding the relationship between human activities, state of \emph{The Art}, and AI algorithms.

\end{abstract}

%% file: contents/1_sotas/c1s1.tex
\begin{quote}
    \small\itshape

    ``To live is to imagine ourselves in the future, and there we inevitably arrive.
    Yet our place in said future may not be the one we envision.''

    \par\vspace{0.3em}
    \hfill\textup{--- Kobo Abe, \emph{In Search of the Present}}
\end{quote}

\section{Human Trajectory Prediction and State-of-the-art Approaches}
\label{sec1}

Human beings always live and grow through their imaginations of the future.
From Oedipus, who learns of a prophesied future and seeks to resist his fate only to bring about that very destiny, to early modern attempts to understand and forecast the world via the Mathematical Principles of Natural Philosophy, people have always longed to know the future and their purported destinies.
Whether seeking to welcome the future or to change its course, this vision of the future has undeniably contributed to human progress and technological development.
From early Greek mythology, astrology, tarot cards, and the Bagua, to predictive machines, time-travel machines, and even the ultimate answers to the universe depicted in works of art, this human longing has existed and continued to evolve in various forms, across various periods.

With the rapid development of machine learning (ML) and artificial intelligence (AI) algorithms, some of the prediction tasks become even more reachable.
They predict the weather, typhoons, stock market trends, and even the movements of celestial bodies, making events that have once been thought impossible a reality.
However, when it comes to the forecasting of human behaviors themselves, even with the state-of-the-art modern technologies and tools, such predictions remain challenging, and progress in these fields continues to be slow.
The reason is quite simple.
Human subjectivity and the uncertainty of human behavior make it even more difficult to encode contexts that are already hard to capture and represent.
Under these circumstances, regardless of the amount of training data available in the context of modern machine learning, most human prediction algorithms tend to learn the data distribution itself rather than the potential relationships between human subjectivity and their future behaviors.

Among several specific forecasting tasks, like the forecasting of human motions, human behaviors, and human intentions, the prediction of human trajectories seems like the easiest one to implement.
It simplifies each person into a 2D or 3D coordinate point, and predicts the future movements of each target in the near future by considering potential interactions between the target person and their environment over a period of time, whether other objects in the scene or other neighboring persons \citep{helbing1995social,alahi2016social}.
Nevertheless, this task, which has often been regarded as a basic computer vision task, remains challenging after even more than 30 years of development with the countless efforts made by various researchers.
In applications such as autonomous driving, it still mostly serves only as a technical aid rather than the core component or the basis of the technology.

As researchers of the human trajectory prediction task, we must acknowledge that this is a direct question we have to face, \IE, what exactly makes this task so difficult.
In contrast, several other tasks that have begun to be developed around the same time, such as understanding and predicting human languages (which forms the foundation of the current large language models (LLMs)) have advanced rapidly.
Interestingly, when comparing these two tasks, it can be observed that their biggest difference lies in the inherent differences in the data they need to process.
When forecasting natural languages, each basic unit they process carries specific human background knowledge.
Every symbol, every word, and every sentence attempts to express some aspect of the human condition or some human desire and hope.
The meanings of these symbols are clear.
They are assigned and assembled by humans.
Meanwhile, these models are indeed capable of capturing the transition probabilities between each of these basic linguistic units, summarized as their ability to ``express'' logic in a manner similar to humans.

On the contrary, for trajectory prediction (including broader human prediction) algorithms, the model only needs to process time series composed of continuously varying numbers, rather than the basic semantic units formed by a finite number of language tokens.
For example, the answer to some specific language-based question could be highly deterministic (or deterministic among several limited choices), especially due to the current question-answer-pair-based training and optimization strategies.
In contrast, humans' future choices are more likely to be \emph{unlimited} (under basic constraints), due to either their varieties within future choices or their unique preferences for perceiving and interacting with the world, formed along with their exclusive upbringing and ways of thinking.
All these properties and probabilities are indirectly and implicitly embedded within several coordinate sequences.
Therefore, teaching machines or algorithms to learn and to uncover these human feelings, interactions, and possibilities of the future from several cold observation sequences has always been the core consideration of human prediction tasks, not limited to human trajectory prediction.

Are researchers really exploring the future of humanity along these lines?
We think the answer to this question would be more pessimistic than expected, even when using what appears to be the simplest human trajectory prediction task as the baseline for discussion.
Back to the very beginning, Helbing \ETAL~\citet{helbing1995social} proposes the Social Force model to characterize pedestrian dynamics.
Unlike approaches that use Newton's laws to describe human movements as the movements of particles with no psychological agency, Social Force treats each pedestrian as an entity with unique personal preferences.
Each person is allowed to have their own independent movement styles and interaction biases.
They are allowed to be attracted to different persons or objects, allowed to reject different ones, and are allowed to have their own personal space boundaries.
In Social Force-based trajectory prediction methods \citep{yamaguchi2011who}, human psychological agency is treated directly as a computable research objective.
They calculate each person's unique attraction preferences and predict their possible and encouraged futures, based on their simulated desired velocity.
In research during this period, humans are allowed and encouraged to be the focus of study.
Algorithms acknowledge their differences and diversity, as well as their unique thoughts and desires.

In the mid-2010s, something has changed.
The foundational technologies of modern AI, \IE, deep neural networks and convolutional neural networks, have seen rapid development.
This development has naturally influenced the task of human trajectory prediction.
Taking Social LSTM \citep{alahi2016social} as a representative method, an increasing number of researchers are employing more adaptive fitting and optimization techniques to represent each pedestrian, as well as potential interactions among them, using learned high-dimensional feature vectors.
This is not something bad, since there are indeed mathematical proofs indicating that neural networks have the ability to fit any mathematical function.
However, an interesting phenomenon has begun to emerge.
When considering social interactions among persons, researchers no longer use traditional methods that are designed to account for differences in different persons' attractiveness or intentions.
Instead, the social pooling and its variations \citep{gupta2018social,sadeghian2019sophie} have been widely used to characterize interactions since then.
They assume that interaction states or contexts are always shared among all interaction participants within a specific area.

This assumption of social interactions may indeed be the optimal one in certain scenarios, as the quantitative results indicate.
However, these methods appear to be gradually undermining human-specific considerations and feelings, especially compared to the former Social Force-based traditional ones.
Taking the above Social Pooling as an example, the interactive representation of specific persons is uniformly obtained by pooling the representations of their neighboring pedestrians' (within specific spatial grids) observed historical states.
In other words, each person's unique interaction preferences could have been \emph{smoothed} as a compromise.

Subsequent researchers like Kosaraju \ETAL~\citet{kosaraju2019social} have also noticed this limitation, so they adopt Graph Neural Networks (GNNs) to model these differences in preferences for considering social interactions.
They think that the ``connections'' between different persons are closely related to the state of all involved persons in the given scenario, and the state of human interactions could propagate through the edges of the graph structure as time goes by.
These GNN-based approaches do indeed improve the performance of prediction algorithms.
They remain the most direct choice in the human trajectory prediction task even today.
However, when examined more thoroughly, an even more astonishing fact is that these methods attempt to treat all the vivid living persons in a scene using a completely consistent structure and shared logic for considering their future.
For example, people with unique behavioral preferences and future plans might be simultaneously encoded within the exact same graph structure, without any structural differences or variations of their own.
In other words, more and more prediction algorithms are likely to put strings on all observed persons, although many researchers may not even realize this point.

With the recent development of high-performance AI models, advanced models have been applied to achieve the goal of better forecasting, whether like the above-mentioned LLMs \citep{bae2024can} or diffusion models \citep{li2024bcdiff}.
However, some other researchers seem to have adopted a more aggressive prediction approach.
They can even discard all of the human-derived prior knowledge, including graph-structured social interactions, and instead feed these high-level black-box models dry sequences of coordinates directly to obtain their so-called human predictions.
More and more researchers start focusing on how to improve prediction accuracy, more specifically, to improve the average or final distance between the predicted positions (even with multiple future choices for those generative prediction models) and the only while single true value, rather than exploring how humans would think and feel when wandering through their future, pondering their possibilities.

It is undeniable that some of the newly published methods are indeed state-of-the-art, due to their best or competitive quantitative metrics across datasets.
However, regarding human prediction algorithms themselves (specifically, a significant portion of which remain inexplicable black boxes), whether these methods can offer humans any new insights or inspire visions of the future remains a question worthy of consideration by our researchers.
Perhaps in a few years, these methods will no longer be state-of-the-art.
So, what kind of resonance can these studies create in our minds?
Can we still say that our prediction algorithms can peek into the predictable future?

%% file: contents/2_art/c2s1.tex
\section{From State-of-the-Art to State of THE ART}

To answer the above questions, as researchers, it could be necessary for us to rethink why we have started to explore and develop human prediction algorithms like trajectory prediction.

\subsection{Human Prediction Algorithms and The ART}

Current AI-based human prediction algorithms are basically considered as part of computer science or engineering technologies in most academic or application classification systems.
Interestingly, in ancient Greek, the word \emph{techne} ($\tau\acute{\varepsilon}\chi\nu\eta$), as the etymological root of \emph{technology}, can refer not only to art, poetics, and intuitive creation, but also to craft, technique, and even the ability to foresee the future.
However, based on the developments in relevant algorithms over the past decade described above, the elements of art, poetics, and intuitive creativity are gradually disappearing from trajectory prediction algorithms.
As a human-centered task, this may not actually be a good sign.
On the other hand, perhaps we can ask another question: Do modern state-of-the-art AI-based human prediction algorithms really not need human intuitive creativity and artistry?

Before starting, we have to indicate that the scope of human creativity and art is actually quite broad.
When discussing these topics, we also need to take into account the properties of these prediction algorithms themselves, rather than engaging in vague ponderings over inscriptions on stone tablets from two thousand years ago.
The definition of modern human prediction algorithms is quite comprehensive from an engineering perspective.
Taking trajectory prediction as an example, the goal of the algorithm is to predict a person's trajectory or activities over a future period, based on observed information from a given period (whether human position coordinates, posture, or contextual information about the environment) while considering their potential interactions with each other or the environment.
Regardless of the type of algorithm, it is almost mandatory to take into account \textbf{perception}, \textbf{interaction}, and \textbf{timing} when building any human prediction models.
Interestingly, these aspects happen to point to an emerging art form, \emph{interactive art}.

Interactive art is a core category that lies at the intersection of digital media, artistic expression, and human-computer interaction dynamics.
Murray \citet{murray1997hamlet} describes the interactive agency within this form of art as \emph{the satisfying power to take meaningful action and see the results of our decisions and choices}.
Represented by video games, and in contrast to art forms such as novels and plays, interactive artworks are not static objects that are pre-packaged and intended for one-way contemplation, but rather dynamic systems composed of code, rules, and input-output feedback.
Some scholars argue that authorial intent and expression are diluted by users (players) and rules in interactive art, claiming that ``video games can never be art''.
We will not elaborate further here on whether these works qualify as art.
As Kocager \ETAL~\citet{kocager2026algorithmic} have pointed out, art should by no means be limited to the works of the European Renaissance.
Nevertheless, it is undeniable that interactive art and human prediction algorithms exhibit a profound structural isomorphism, whether due to rule restrictions, participants' freedom of choice, or the author's lack of complete control.
Such isomorphism could be decomposed into these three aspects correspondingly.

(i) \textbf{Rule Restrictions:}
Participants or players participating in interactive art are allowed to present their agency.
This does not mean that they have full control over these artworks.
Instead, their behaviors are under strict rule restrictions established by the artist, whether in terms of the worldview, the reachable areas, or the types of allowed activities.
These requirements also apply to human prediction algorithms.
For example, our considered predictions are always associated with some specific prediction scenes.
Whether described by images or videos, these scenes contain their unique environmental constraints to restrict human behaviors.
In addition, when considering interactions among humans, certain social rules are usually indirectly embedded across these contexts, determining in which way they interact with others.
Some prediction algorithm researchers also refer to this point as \emph{Social Etiquette} \citep{robicquet2016learning}.
The difference is that rules in interactive art are often established by the artist, whereas the rules considered in predictive algorithms are those formed naturally by humans throughout their development process and are described and modeled using specific algorithms.

(ii) \textbf{Freedom of Choice:}
Under the terms of the given rules, players in interactive art are granted a considerable level of freedom and control.
They are allowed to move freely within a specific space, perform certain actions independently, and experience the various consequences resulting from those actions.
This agency is also considered the most significant distinction between interactive art and other forms of art \citep{murray1997hamlet}.
For the prediction scenes considered by human prediction algorithms, their considered prediction targets are also assumed to behave and to interact with others freely, restricted by the given social rules (though they are not mandatory in most scenes).
Most prediction algorithms do not assume that individuals in a given scene behave according to a specific or unique pattern.
Rather, they allow these data to exhibit unique distributions and permit variations among prediction targets.
Based on this consideration of agency, the prediction targets in human prediction algorithms are also widely referred to by researchers as \emph{agents}, which have been also extended to pedestrians, cars or even robots.
From this perspective, these considered agents and players of interactive art exhibit more similarities, and the agency of such agents in prediction algorithms extends even beyond that granted to players in interactive art.

(iii) \textbf{Incomplete Control:}
As one of the compromises for allowing the agency of players or participants of interactive art, artists may not have complete control over or be able to fully determine how players will experience their artistic expressions.
This is even more directly reflected in interactive art set in open worlds.
Before reaching a predetermined key point, each player may have chosen their own unique path through the experience, which then shapes their own distinct artistic experience, naturally leading to different interpretations of the artwork.
This lack of complete control is reflected more directly in human prediction algorithms.
The algorithm itself has no authority and no ability to shape the future trajectory and activities of the prediction target.
The target is also assumed to exhibit multimodal preferences in the future, rather than being confined to a single, predetermined future (though this remains technically questionable from the perspective of current evaluation metrics).
This is also a major reason why prediction algorithms have hit a bottleneck in recent years, as they lack clear and objective evaluation metrics unlike other deterministic AI tasks (such as image recognition), just like it is difficult to evaluate whether a player has \emph{correctly} experienced interactive art or not.

\NEWHALFLINE
These structural similarities, especially given that both interactive art and prediction algorithms are human-centered, could have in fact provided considerable insights for the development of algorithms themselves.
Unfortunately, even today in 2026, most relevant research still focuses on how to assemble their cutting-edge AI models, combined with unique optimization strategies in order to demonstrate that their methods can achieve the best quantitative prediction performance.
Most of these methods do not, or are not even encouraged to, explore which mechanisms are actually effective to help model which part of the prediction algorithm beyond mere numeric improvements.
Only a few researchers have made several attempts to dissect the black-box networks layer by layer, to investigate whether AI models have learned or uncovered insights about human behavioral patterns or interaction habits that are difficult for humans to recognize but might be revealed by AI models.

In recent publications, Wong \ETAL~\citet{wong2023socialcircle,wong2024socialcircle+,wong2024resonance,wong2025reverberation} have built a series of human trajectory prediction algorithms, named \SCMODEL, \REMODEL, and \REVMODEL.
Unlike other methods, these papers devote more efforts to analyzing exactly how the models process human perception, human interaction, and human temporal responses to events, respectively.
In the appendix to \REVMODEL~\citep{wong2025reverberation}, the authors even mention that these methods are conceived as a trilogy:
\begin{quotation}
    ...
    Through this trilogy, the focus of our research evolves from where interactions happen (\SCMODEL), to how they co-vibrate across time (\REMODEL), and finally to when and for how long their influences live (\REVMODEL), pondering the question: how perception, causality, and anticipation are intertwined.
\end{quotation}
This actually provides us with a rich source of material for exploring the structure of AI prediction algorithms in relation to human behavior itself, as well as the nature of human interactions revealed by AI prediction models (for example, Wong \ETAL~\citet{wong2024resonance} discovers specific time-frequency responses in the trajectory components of human interactions with different intentions during the experiment).
It is particularly worth noting that, although other researchers like Zhang \ETAL~\citet{zhang2019sr,zhang2023towards} or Bae \ETAL~\citet{bae2024can,bae2025social} have also attempted to analyze these topics from different human perspectives in two consecutive papers, we find that very few researchers have published their findings in a series of publications guided by the same perspective for understanding human behaviors (perhaps it is the norm in the field of human prediction for researchers to tackle a brand-new topic each year or even to follow in others' footsteps in every paper).

What is even more interesting is that, based on the materials these authors have already published and provided, the influence of interactive art on these works is evident, particularly like video games \emph{Death Stranding} (2019), \emph{Death Stranding 2: On The Beach} (2025), and \emph{The Last Of Us Part II} (2020) published in recent years.
In the following sections, we will attempt to use the perspective of interactive art to analyze how the authors perceive human motions, how they incorporate human feelings into AI-based prediction algorithms, and finally how these human perceptions function within the AI models.
Our goal for these discussions is also to provide new perspectives for understanding human prediction algorithms with a more human-centric point of view beyond mere numeric gains, especially taking into account these artworks that carry the distinct marks of the post-COVID era and have been published around the same period as these algorithms.

%% file: contents/2_art/c2s2.tex
\subsection{Perception: Connections}

As we have discussed above, it is almost mandatory to take into account perception, interaction, and timing when building any human prediction models.
We first concentrate on the perception stage, attempting to re-understand and re-frame the modeling of human perceptions from the interactive art perspective.

Simply put, in this stage, prediction algorithms gather all observed information for each unique prediction target, providing basic representations for making further forecasts.
During this stage, from the perspective of computer science or computer vision specifically, it is quite clear what the model needs to do, \IE, simply construct preliminary contextual representations for each agent (like pedestrians, cars, etc.).
As we have introduced in \SECTION{sec1}, over the past decades, researchers have used manual methods such as Social Force \citep{helbing1995social} to carefully construct and calculate preference parameters for each agent, as well as Social Pooling \citep{alahi2016social} to construct shared state representations end-to-end.
In recent years, they have also experimented with advanced models such as graph structures and LLMs to further refine and represent these contexts.
From an engineering and technical perspective, these methods appear to simply use different mathematical tools to achieve the goal of constructing specific representations for describing how humans perceive during the (historical) observation period.
However, the considerations regarding how the persons being predicted are modeled to perceive the world and to connect to others are clearly different in each of these methods.

As for interactive art, it is interesting that the period of rapid growth in interactive art happens to coincide with the development of these algorithms mentioned above.
Due to limitations in computer hardware, the forms of interactive art in the early 1990s have been relatively limited.
As represented by \emph{Super Mario World} (1990) featuring side-scrolling gameplay, and \emph{The Legend of Zelda: A Link to the Past} (1991) with its top-down perspective, artists still rely on relatively simple methods to present worlds to players, providing the necessary context through a relatively fixed third-person camera angle in a fairly limited manner.
In these works, the artistic experience perceived by different players may be nearly identical.
With advances in computer graphics technology and computing capabilities, platforms supporting full 3D technology like PlayStation (1994) have begun to appear, offering artists new alternatives for considering how players experience their worlds.
For example, the artists in \emph{Resident Evil} (1996) utilize differences in perspective and shifts in viewpoint to offer players a new way to experience the game's horror atmosphere.
\emph{Super Mario 64} (1996) even allows players to freely move and rotate the 3D camera, enabling every player to perceive the world in their own unique way and enjoy a distinctive artistic experience.

Interestingly, even in human trajectory prediction algorithms for the year 2026, a significant number of researchers still model how predicted agents perceive the world in ways that remain at levels comparable to the interactive art of that era.
For example, a significant number of researchers still rely solely on 2D displacement vectors (sometimes even after gridding) between coordinates in the 2D space within a unified coordinate system to represent the positional relationships between the target agent and other scene objects or other agents, encoding these as contextual representations.
It is worth noting that we are not criticizing these methods for not structuring their networks in the same way as other art forms, but rather discussing the possibilities and similarities between interactive art and human prediction algorithms.
From this perspective, both actually share very similar starting points within their developments.

From the 2000s to the 2010s, interactive art has once again undergone a significant revolution and evolution, benefiting from further advancements in computer graphics technologies.
The most significant difference is that artists have begun to place greater value on narrative as an essential component of interactive art, particularly by deeply integrating cinematic, immersive narratives with interactive and perception experiences.
Just as in cinematic and television arts, artists now have more space to explore the inner lives of characters, offering players an immersive experience that reveals how different characters perceive the world in their own unique ways.
From the early \emph{Grand Theft Auto: Episodes from Liberty City} (2009) and \emph{Grand Theft Auto V} (2013) to the later \emph{Heavy Rain} (2010) and \emph{Detroit: Become Human} (2018), artists have further incorporated shifts between multiple character perspectives to create their unique artistic narratives.
Here, players can gain a more intuitive understanding of how characters with different backgrounds and different states perceive the world in different ways.
For example, in \emph{Detroit: Become Human}, when characters scan their surroundings to perceive the world initially, the background is covered with clean 3D grid lines.
However, these grid lines will become distorted once the characters trigger a specific event and break the rules, reflecting the inner changes taking place within these characters.

Interestingly, among other human trajectory prediction algorithms from the same period, a few researchers like Xie \ETAL~\citet{xie2017learning,barata2021sparse} have also attempted to describe how agents within a scene perceive their environment and context by constructing and then simulating how agents would ``scan and perceive'' a special ``field'' that exists within the scene.
Although we cannot be certain of the authors' true motivations for designing these elements, it is still fascinating that the period when these algorithms have become popular coincides remarkably with the emergence of the mentioned cinematic narratives in interactive art, characterized by shifts between multiple main characters.

An interesting turning point has occurred around the year 2020 to the mid-2020s.
For interactive art, during this period, artistic expression is no longer constrained by hardware limitations, but rather by the artists' own visions for their virtual worlds.
Particularly noteworthy is \emph{Death Stranding}, released in late 2019, which has received quite poor initial reviews from players.
Some media even describe it as a merely passable, unremarkable work, arguing that the artist's artistic expression is to some extent a form of ``pretentiousness''.
Interestingly, after players have genuinely experienced those years of global emergency events, the evaluation of this work seems to have undergone a dramatic reversal, even though the content and artistic perspective of the work itself have remained mostly unchanged.
The developers have also repeatedly expressed their artistic vision, which centers on the exploration of connections between people.
In fact, it is after experiencing traumatic events in the real world that players gradually form a new understanding of this concept.

It could be quite difficult to directly discuss this concept of the abstract connection.
In the \emph{Death Stranding} game, the world has undergone a massive disaster.
People are forced to take shelter underground, cut off from one another.
The player takes on the role of a delivery person tasked with transporting essential supplies between different shelters.
The story unfolds as he travels across the continent, connecting the isolated persons.
Here, the equipment named Odradek is used as the terrain scanner in the beginning of the game.
In the game world, it scans the front nearby fan-shaped areas by producing some kind of specific sound waves.
Holographic markers will be temporarily displayed in the world once the scanner's signal encounters different environmental objects (footprints, items, roads, or rivers) or other entities (like other porters, enemies, etc.).
After scanning, safe areas like roads will be marked as references for players to decide their future paths, and other interaction targets like cargo and entities will also be highlighted.
Odradek is also used by other porters or enemies.
They also use their Odradeks to scan terrain and cargo, and to detect any other potential risks.
During such a process, no more human words or languages are needed, only those scans, just like marine animals' \emph{echolocation}.
They use their Odradeks to scan the environment, to plan routes, to interact with others, and finally to connect the broken world.
In the late 2010s, with abundant resources and superior living conditions, players have even found it difficult to understand this setting or to grasp why such connections are necessary.
It is exactly this difference in experience between the player and the character they play that leads to the different evaluations of such artworks.

In particular, connection is also a crucial aspect that researchers of human prediction algorithms cannot overlook, especially for simulating how agents would perceive their surroundings and the world.
Notably, Wong \ETAL~\citet{wong2023socialcircle} actually incorporate this concept of Odradek-scanning into their human trajectory prediction algorithm.
The authors also explain on their website the influence that the game \emph{Death Stranding} has had on their development of this model.
Specifically, in their proposed \SCMODEL, they no longer use previously widely employed relative distance vectors to describe the relative positions of different agents.
Instead, they employ a circular scanning mechanism similar to that of Odradeks.
Each predicted agent is considered as the center of its own unique circle.
When describing its observed context, they utilize different angular partitions as indices.
Similar to echolocation, they use angles to determine the interaction context in different directions, where partitions containing other pedestrians that are closer and moving faster are assigned a higher feature contribution, just like how Odradeks react to terrains with higher risks like the rapid-flowing rivers for the players to plan their future behaviors.

This type of AI prediction algorithm design, which is almost homomorphic to conceptual design in modern interactive art, is quite rare in this field.
In addition to these structural designs, Wong \ETAL~\citet{wong2023socialcircle} even discover some phenomena in their experiments that other models struggle to uncover.
For example, when analyzing attention scores for angular partitions, if a neighbor is traveling alongside the agent, the model assigns a higher attention weight to the partition representing that companion than to the partition representing the agent itself.
Furthermore, within the predicted trajectories, they consciously maintain their relative positions within the group and avoid crossing their companions' paths.
This phenomenon is quite intuitive.
As pedestrians walking along the street, we might find it difficult to systematically list exactly what we are paying attention to.
However, within this model architecture, the network is actually able to extract some ``human-like'' emotional aspects of pedestrians from the training data.
Although such conclusions might seem obvious, these results are derived through reverse deduction based on the algorithm's patterns.
This may represent a brand-new approach to understanding the agency of humans itself, even though it is actually caused by the small design to simulate how humans perceive the world.

%% file: contents/2_art/c2s3.tex
\subsection{Interaction: Vibrations}

From the above discussions we mention that the series of trajectory prediction algorithms by Wong \ETAL~\citet{wong2023socialcircle,wong2024socialcircle+,wong2024resonance,wong2025reverberation} follow the same core metaphor of echolocation of marine animals.
From the interactive art point of view, marine animals are significant characters in the two \emph{Death Stranding} games.
From stranded whales, the beach, to the giant beached things (BT) enemies like dolphins, octopuses, stingrays, and even shore crabs, marine animals serve as symbols of the transition from life to death, as well as the returning from the world of the dead in the game world.
Their communications are considered as the concept of ``connections'' themselves.
Such concepts have been even strengthened in its sequel, \emph{Death Stranding 2: On The Beach} (2025), where the ocean, the beach, and those marine animals all play significant roles in advancing the plot.
We first attempt to analyze the meaning of marine animal imagery as interpreted through the perspective of artists.

Of all these marine creatures, whales appear most frequently.
In both games, safe house or base of operations for the player's character, the distribution center as well as the porter entrances of knot cities, are all shaped like whales.
Specifically, based on the published art books, these distribution centers might be initially imagined as whales that have just surfaced from the sea.
In the original design, the antenna on its head resembles a whale spouting water.
It is also quite interesting that for large baleen whales, their spouts can reach vertical heights from several to over ten meters, making them visible to conspecifics from several kilometers away.
When traveling at the sea surface, loosely dispersed pods or mother-calf pairs often observe the positions and timing intervals of each other's blows to maintain formation and synchronize their movement without producing underwater sounds, thereby avoiding detection by predators like killer whales.
This way of communication is quite unique while easy to understand.
It is similar to the modern communication technologies, where signals will be broadcast through the amplitudes and the frequencies of some other carrier signals.
Different frequencies carry different information, while information transmitted at the same frequency represents the same communication content.
Although the artists do not explicitly point it out, the term ``resonance'' could just describe such communication preference, as well as the wider interactions among persons.
Here, the artists have artistically interpreted their understanding of human social connections and interactions through the element of whale echolocation and communications, allowing players from diverse backgrounds to experience the essence of this connection, especially in the current post-COVID era.

Back to the prediction algorithms' side, if scanning is the very first step for locating and representing environmental objects, the simulating and modeling of interactive behaviors (simply referred to as interactions) among different agents to be forecasted will then become its another significant but subsequent step.
In this process, researchers need to design specific structures for aggregating interaction information to enable AI models to sense and understand how human interactive behaviors emerge and evolve.
It is worth noting that the complexity of human interactions makes it difficult to describe and simulate the impact of human interactions on their future trajectories or behaviors through simple, individual mathematical operators or equations.
Under these circumstances, researchers have employed various approaches to construct numerous unique sets of equations or computational models to attempt to simulate this human behavior.

Whether repulsive and attractive potentials in Social Force \citep{helbing1995social}, grid-based spatial pooling features in SocialLSTM \citep{alahi2016social}, or recent prompt templates in the natural language form \citep{bae2024can}, researchers have demonstrated considerable enthusiasm and dedication in their efforts to find different mathematical tools to describe these unstructured patterns of human interactive behaviors.
We do not deny the efforts of these researchers.
On the contrary, we greatly appreciate their creativity and curiosity.
However, it cannot be overlooked that most researchers still treat the improvement in the quantitative prediction performance of AI models as the sole criterion for evaluating the effectiveness of these interaction-modeling components.
Under this evaluation system, researchers may not be encouraged to deepen their comprehension of human behaviors themselves, particularly in publications with strict page limits.
This creates a strange model ``illusion'', making it difficult even for the researchers themselves to determine whether their consideration of these interactions can truly guide their model in recognizing human interactions.

However, for interactive art, the concept of interaction itself is actually the core consideration of most artists.
Whether it is human-computer interaction (how human actions are conveyed to the computer and how feedback is received), game AI\footnote{
    Game artificial intelligence is also known as game AI.
    However, the term ``AI'' here is not the same as the term AI that is currently widely used to refer to machine learning or other computer technologies or algorithms.
} (interactions between the player's character and other non-player characters (NPCs)), or social interactions among players (interactions between the characters players control), complex interactive systems serve as the set of basis vectors for constructing the entire interactive artwork.
The role of social interactions considered within human prediction algorithms is similar to a hybrid of the above game AI and social interaction among players, where the latter ones showcase greater variety while reflecting unique artistic vision of the artists.

Among these interactions, particularly those between players, the asynchronous online interaction systems are especially noteworthy, like those in ``Souls-like'' games, such as the \emph{Dark Souls} series (2011 - 2016) and \emph{Elden Ring} (2022), as well as in the \emph{Death Stranding} series (2019 - 2025).
These interactions do not rely on real-time actions or information exchanges.
Rather, they serve to express the players' ``spectrum of emotions and needs'' within specific contexts and areas.
For example, in these games, players are allowed to place template-based messages, signs, different types of buildings that could enhance the gameplay, or even to share their footprints or trajectories with others.
These asynchronously updated messages or buildings are triggered when a player's character reaches a specific location.
They directly convey to the current player the emotions of others who have been there before.
Whether it is the ``I did it!'' after overcoming a difficult battle, a brand-new bridge over a raging river, or even a ``jump down'' at the entrance to a secret passage (if that really were an entrance), these elements often resonate deeply with players, even if they have never met in the same time and space.
Combining the imagery of whales in interactive art mentioned above, along with their biological instinct to communicate using carriers of specific frequencies, an interesting fact is that the artists have further abstracted the concept of interaction between players.
These interactions need not occur at the same time, or even in the same place.
They only need to convey some unique emotional frequency bands within the broad spectrum, belonging and resonating exclusively to the participants of the specific interaction.
This is the unique interpretation that artists offer in interactive art, drawing on their special understandings of human interactions.

Wong \ETAL~\citet{wong2024resonance} have also noted that the interactive imagery in these games, as well as some other forms of art (like they also mention the TV series \emph{Severance} (2022 - 2025), whose central concept involves signals of different frequencies triggering different personalities of characters), prompt them to use the concept of ``resonance'' to directly model social interactions in their human trajectory prediction algorithm \REMODEL.
Unlike other methods that represent interactions in terms of time or space, they first use Fourier transform or wavelet transforms to analyze trajectories from the additional spectral perspective.
Then, the final interaction representations are constructed by calculating the spectral similarity between different interaction participants.
This thought is quite straightforward.
The authors even do not provide any additional human prior, yet they have discovered interesting experimental phenomena.
Specifically, their AI prediction algorithm has actually extracted an experimental finding from the data indicating that people whose trajectories have similar spectral distributions are more likely to walk together or to interact with each other.
This coincides with what the common expression ``on a similar spectrum'' has indicated in human interaction scenes, demonstrating the AI model's unique perspective on understanding human interactions even from trajectories.

In addition, the authors also present some additional experimental results regarding how AI models understand human interactions from the perspective of vibrations.
For example, social-sourced vibrations (vibrations in trajectories exhibited to handle interactions) exhibit more nonlinearities.
Furthermore, they also mention that self-sourced vibrations (vibrations in trajectories aimed at achieving one's own intentions) and social-sourced vibrations vibrate almost vertically in space.
Similar to our previous conclusions, these aspects are likely difficult for us humans to notice during interactions, and most other prediction methods have failed to capture these phenomena.
This also points to a rather interesting way of understanding humanity.
Starting from artists' insights into human nature, inspiring the design of AI models, which in turn reveal some fascinating patterns.
Although the definitive confirmation of these phenomena still requires extensive experimentation and field research, these findings offer new entry points for our exploration of human thought, as well as our subconscious understandings of human interactions and curiosity of the future.

%% file: contents/2_art/c2s4.tex
\subsection{Timing: Latencies}

After the modeling of how humans perceive their observed contexts and understand their potential future interaction choices, another key consideration in human prediction algorithms is how to represent and map these perceptions and interactions onto different temporal indices.
For example, they need to consider which moment in the observation sequence is more important for future decisions, and which past moment should be weighted more when forecasting the next moment's trajectory, motion, or behavior.
This is a fairly important aspect of prediction algorithms, where researchers have also devoted a significant amount of effort.

In traditional machine learning-based prediction algorithms, researchers use probabilistic models such as Bayesian and Markov models to analyze the timing and transition relationships between sequences.
These works (like work \citep{kitani2012activity}) typically involve rigorous mathematical discussions based on probability, aiming to capture the inherent properties of sequential data.
Subsequent methods based on recurrent neural networks (RNNs), including the above-mentioned Long Short-Term Memory (LSTM) based SocialLSTM \citep{alahi2016social}, constitute a natural extension of this idea into the era of deep neural networks, learning the relationship between the next prediction time step and the current sequence state, simulating the state transition patterns in the time series.
With the introduction of sequential attention and the corresponding Transformer network \citep{vaswani2017attention}, researchers have adopted a more unified approach to analyzing and predicting time series, by using attention algorithms to determine which parts of the input should be assigned higher weights, and then utilizing these weighted representations of the sequence to predict future outcomes.
As a result, in current algorithms for predicting human trajectories, researchers mostly concentrate on the construction of interaction representations, and then ``mechanically'' feed them into a Transformer or similar backbone to learn sequence interpolations and extrapolated states.
The timing appears to be the part that places the highest demands on mathematics and theoretical proofs, while the gap between it and human experience and perceptions of the future is enormous.
Consequently, very few researchers seem to have directly considered this problem from the human perspective, especially from how humans construct their future visions.

Notably, from the interactive art side, artists have actually demonstrated a considerable interest in presenting how they feel the concept of time itself, especially considering that the timeline arrangement is another unique but major concern when constructing any interactive artworks.
From the perspective of the applicable time scales, this timeline arrangement can be considered to comprise two entirely distinct levels: the frame-level timing used to manage fine-grained character control or behavior, and the narrative chronology used to manage the overall interactive narrative or artistic expression.
Interestingly, to some extent, research on timing in human prediction algorithms lies exactly at the crossover point between these two types of timeline management.
On the one hand, when dealing with specific, detailed interactions, people's planning for future activities exhibits quite precise frame-level differences, such as sprinting to dodge an oncoming out-of-control car.
On the other hand, people's planning for the future can itself be viewed as a forward-looking prediction of the future.
They write New Year's resolutions, work plans, and other documents to manage their future expectations on a macro level.
At this scale, human prediction algorithms might have to simultaneously take into account a mixture of both aspects.
Before further discussions, we will first introduce these two scales of timing in interactive art, as well as whether such considerations or requirements could also be taken into account in human prediction algorithms.

Compared to other forms of artistic expression, frame-level timing is actually an additional requirement unique to interactive art.
It directly determines the effectiveness of the ``interactivity''.
For example, it specifies those uniquely-shaped temporal rules like in which specific frames user commands should be accepted, and in which frames interactions with other objects should be further carried out or dropped directly.
In early action arcade games, such as \emph{Mortal Kombat} (1992), this kind of timing is quite straightforward.
Corresponding to the refresh rate of the display device, the rendering of each frame and the corresponding interaction logic processing are directly tied to the hardware screen refresh (like 60 frames per second).
Within each of such frames, the outcome (or called the \emph{response}) of the program to every player input is described by the states of a temporally continuous finite-state machine (FSM).
For example, after the player triggers the ``Uppercut'' in \emph{Mortal Kombat}, the controlled character will be temporarily locked for 7 frames while playing the start up animation, turning into the ``startup'' state while showing the player and the opponent that an attack is about to happen.
After these frames, the state becomes ``active'' and then holds for another 6 frames, animating the attacking frames of the character to indicate this ongoing interactive behavior while implementing the attack logic and damage calculations in the background.
The state finally turns into ``recovery'' after these 6 attacking frames, keeping for another 25 frames for cooling down the entire attack behavior, meanwhile completing the iteration through all states of the FSM.
Interestingly, no matter how times have changed, even until now, most games still use FSMs or their various variations to provide players with distinctive senses of impact of how they feel and interact within the frame-level timeline (or interactive feedback more broadly) in interactive art.
For instance, this ranges from \emph{Sekiro: Shadows Die Twice} (2019), which offers players satisfaction through Deflect (precise blocks) that require precise timing for handling enemies' attacks, to \emph{Monster Hunter Wilds} (2025), which delivers a unique experience through combos and immediate combat feedback.

FSM is not something new from mathematical analyses or the human prediction algorithms' perspective.
For example, when modeling social interactions among pedestrians, Yan \ETAL~\citet{yan2014modeling} categorize pedestrians' interactions into six states, including walking together (Walk), passing (Pass), spreading (Spread), following resulting in walking together (Follow-Walk), following resulting in passing when the followed subject is slower than the other (Follow-Pass), and following resulting in an increasing separation between subjects when the followed subject is faster than the other (Follow-Lost).
This is a rather interesting early attempt.
It not only establishes several discrete states but also attempts to define transition relationships between some of these states at the observation frame (by packing them as three other individual states), such as transitions from the Follow state to other states like Walk, Pass, and Lost.
Although this method does not actually define a complete FSM, \IE, neither having complete traversal conditions nor complete state transition equations or matrices, this actually demonstrates once again the different yet similar interpretations of timing in human interactions within human prediction algorithms and interactive art.

In addition, another major difference between the above frame-level timing and that attempt within prediction algorithm is that the temporal scale of each interaction state has not been explicitly defined, like under what conditions that a specific state will turn into another one.
In this context, the state is more like a classification result generated frame by frame rather than a structural representation of human interactions based on specific states and their transitions.
However, it is worth noting that the diversity of human behavior (even when specifically referring to the interactive behaviors that prediction algorithms focus on) makes it difficult to enumerate or generalize all human motions or choices using limited hard labels, not to mention some accurate numerical temporal latencies or delays of humans to further handle these state-changes.
From the perspective of 2026, there is almost no method left for describing human interactions using such labels or state-transition relationships, except for those several natural language-based prediction algorithms like proposed by Bae \ETAL~\citet{bae2024can}.
For example, they use question-answer templates like ``\emph{Question}: With which pedestrians does pedestrian $\{n\}$ form a group? \emph{Answer(s)}: Pedestrian $\{n\}$ forms a group with pedestrian $\{k\}$. / Pedestrian $\{n\}$ will walk alone.'' to tell the interaction state to LLMs for making further predictions.
Although these approaches do demonstrate excellent prediction performance, our concerns regarding frame-level timing have still not been directly taken into account.
Even though it might be quite intuitive that we could unconsciously make arrangements (whether rough or detailed) along the timeline when interacting with others or planning our future paths, unfortunately, very few researchers have noticed this rather intuitive human perception within their human prediction algorithms.

In these situations, it becomes quite noteworthy that the \REVMODEL~\citep{wong2025reverberation} have conducted detailed experimental analyses at the exact frame-level, such as ``it considers that event occurring at the historical step 1 will be postponed for about 6 steps (4.8s) to allow the ego agent to handle it, ... After that, such event will receive the highest concentration for about 2 steps (1.6s) to decide future trajectories''.
This exhibits a remarkable similarity to the transitions between states in action games we discussed earlier.
The authors even mention on their website that the narrative chronology in some interactive artworks, like \emph{The Last Of Us Part II} (2020) and \emph{Death Stranding 2: On The Beach} (2025), also contributed to the development of their model.
This raises a rather interesting point that lies precisely at the intersection of the two approaches to handling timing or timeline arrangement in interactive art.
Before further discussing their model details or experimental findings, we first talk about the narrative chronology from the art perspective.

Different from the above discussed frame-level timing, the narrative chronology is not unique to interactive art.
Instead, the diverse understandings or interpretations of the concept or the properties of time itself could reveal artists' fascinating perspectives on timing within the sequence of human life, even regardless of the form of artistic expression.
The narrative chronology is actually intended to distinguish between the story's factual ``story time'' (such as when certain events occur) and the ``narrative time'' conceived by the artists and presented to the readers or audiences (like the selectivity in the order in which different events appear or play).
Specifically, artists working in film or theater need to utilize specific editing and narrative strategies (like cross-cutting, flashbacks, and flash-forwards) to enable the audiences to understand the background, progression, and consequences of the presented events within a limited period of time, as shaped by the artists' unique preferences.
This provides different perspectives from which audiences perceive and understand events.
For example, in \emph{Memento} (2000), the sequence of all events experienced by the main character has been rearranged along two narrative timelines that grow in opposite directions.
Events from two timelines are cut together in a cross-cutting montage and converge at the end of the film, presenting the audiences with the male protagonist's unique experience of being unable to judge these cause-and-effect relationships as a result of his short-term memory loss.
This distinction between the story timeline and the narrative timeline has also contributed to the artistic vision behind \emph{TENET} (2020).
Its narrative timeline develops naturally, but the story timeline incorporates more changeable events.
Here, the passage of certain events along time will be assigned a coefficient of -1 in some cases, causing these events to exhibit a unique reverse chronology, providing novel watching experiences for the audiences.

When distinguishing the story timeline and the narrative timeline, interactive art deepens the interactive experience, allowing artists and players to explore more options for immersively experiencing the unique artistic sensations.
For example, in \emph{Beyond: Two Souls} (2013), the narrative timeline is completely disrupted.
The artists enable players to experience the protagonist's various fragments of her growth at different stages within her life, allowing them to empathize with her unique experiences and traumas.
Artists of \emph{The Last Of Us Part II} (2020) further strengthens the differences between these two types of timelines.
Among the two main characters in the story, players must simultaneously experience four unique story timelines (including Ellie's present and past, as well as Abby's present and past, which are completely symmetrical), all of which are intertwined at different points in the narrative timeline.
In some scene, the player might be attacking another protagonist who serves as an enemy, while in the next scene, the player might then experience some small daily moments of that hostile protagonist's past.
This narrative chronology offers players an entirely novel experience, particularly by allowing them to immerse themselves in the temporal latency between the occurrence of events and the consequences they finally bring.
This dislocated narrative experience is something that other art forms might hardly provide.
The extensive discussion that followed this artwork's release further validates this point, especially for providing players different perspectives for perceiving their own self-interpreted gameplays.

Although there is a significant difference in the formal approach between the timing considered in human prediction algorithms and this kind of narrative chronology, as a human-centered task, people's perceptions of different events, as well as the trajectories, actions, and behaviors they actually carry out, correspond precisely to the story timeline and narrative timeline described above.
Specifically, humans will naturally imagine future expectations in their minds before performing specific actions, just as described as ``prefactual thoughts'' \citep{epstude2016prefactual}.
This means that what human prediction algorithms mainly considers is actually what they have actually done, rather than what they have thought or planned in their minds.
Although some researchers have attempted to adopt a human-centered perspective by simplifying human intentions into either the endpoint or waypoints along a future path \citep{mangalam2020not} from this point, like the idea that \emph{It is not the journey but the destination}, there are still very few methods for understanding future temporal plannings from human perspectives, even though this topic arises directly from our intuitive imaginations of the future.

Similar to the above narrative chronology within interactive artworks, Wong \ETAL~\citet{wong2025reverberation} propose a rather interesting thought for considering the planning of future trajectories by introducing the concept of latencies of trajectory-changing events as well as its corresponding ``rehearsal domain''.
Like how story and narrative timelines have been differentiated in interactive art or other art forms, their proposed \REVMODEL~provides a new system-theory-like way to reinterpret how humans plan their future interactions and trajectories.
The entire implementation process simulates temporal properties (\IE, their defined trajectory latencies) of how humans react to different trajectory-changing events, by learning how long these events will take to start modifying their future trajectories or behaviors, how long these modifications will last, and finally how long these influences vanish in the future.
These learned latency modes will then be used to compute future trajectory representations as their imagined future, which will be finally refined and processed to become the final model predictions.
It is also quite noteworthy that this approach takes both our previously discussed frame-level timing (like how long some state may persist) and narrative chronology (differences between different timelines) into account when forecasting human trajectories.
In their published materials, the authors state that interactive art such as \emph{The Last Of Us Part II} has played a significant role in the conceptualization of their model.
They have provided detailed analyses of the artistic insights they gained from the narratives and interweaving of different timelines in both \emph{The Last Of Us Part II} and \emph{Death Stranding 2: On The Beach} (we have omitted them for space reasons).
At a time when most human prediction algorithms still focus or are encouraged to only focus on a single quantitative metric of performance, this approach that draws on human experience is quite intriguing.

Further, like the above mentioned two other papers by these authors \citep{wong2023socialcircle,wong2024resonance}, 
experiments in the \REVMODEL~paper also find some interesting data modes in the urban city pedestrian walking scenes.
Simply, they find that most of the peaking latencies of trajectory-changing events (how long it will take to start handling interactive events or changing future trajectories) they computed fall within the specific range of about 1.6 to 6.4 seconds, varying according to the different motion preferences of pedestrians.
Interestingly, this value does not change with variations in the sampling frequency of the observed data, thereby also satisfying the system requirements of the Nyquist-Shannon sampling theorem.
As pedestrians walking down the street, we may find it difficult to consider how specific interactions will be processed precisely after a certain number of seconds after we notice these occurrences.
Consistent with our previous thoughts about those data modes behind how humans perceive and interact, although the further confirmation of such latency data may require extensive field investigations, these unique data reversely deduced by AI prediction algorithms might have the ability to even further inspire some additional understandings or interpretations of ourselves regarding our future.

%% file: contents/3_unpredictable/c3s1.tex
\section{From the Art To the Future}

The human desire for the future has, to a certain extent, consistently driven humanity's long-term development.
This longing has been passed down, documented, and carried forward in various forms, whether they are written symbols on cave walls, parchment, or bamboo slips, analog signals on tape and film, or frozen digital memories buried deep beneath the Arctic ice sheet\footnote{
    The GitHub Arctic Code Vault program (2020) aims at saving all human codes for at least thousands of years under Arctic to prevent any climate change or natural disasters, saving open source codes as cultural heritages of softwares, algorithms, and even further digital artworks.
}.
In recent years, AI has become a hot topic.
However, even the most complex AI systems are composed of modular AI algorithms.
These AI algorithms are tools created by humans to perform specific tasks.
From this perspective, they are very similar to human artistic expression.
People convey their ideas and share their reflections about the world through various forms, such as writing essays, painting oil paintings, or even taking photographs and videos to permanently record their own memorable moments.
To a certain extent, this way of thinking about the world also remains a key component of AI algorithms.
For example, humans need to establish rules for AI algorithms, define the boundaries of commands, and specify execution methods, which just like how they create their artworks.
As the term ``techne'' ($\tau\acute{\varepsilon}\chi\nu\eta$) suggests, technology can be both artworks used to convey cognitions and emotions, and crafts for manufacturing or performing operations.

However, we have observed a strange phenomenon in recent studies.
Even with what appear to be the simplest human trajectory prediction algorithms, fewer and fewer researchers are willing to incorporate their human perspectives into their constructed AI prediction models, particularly their cognitive approaches and their hopes for the future.
Although these methods do indeed achieve excellent performance on certain quantitative metrics, we cannot help but ask: what can these AI prediction algorithms offer us once we stop considering these metrics?
Are there any potential different human thoughts or interpretations from the AI prediction algorithms' point of views?
As we have discussed, we have attempted to reconsider the human-centered aspects of human prediction algorithms through the perspective of a specific art form, \IE, interactive art, whether it involves the simulating of how humans perceive the world, how they think about interactions, or how they approach the management of timings.
What is particularly interesting is that we have found similarities between human prediction algorithms and interactive art in all three of these aspects, especially in the similar approaches taken by researchers and artists in addressing similar challenges from different fields and with different backgrounds.
Some interactive art-inspired human prediction algorithms can even identify and summarize underlying patterns of human behaviors from data, which also offers new perspectives for approaching AI algorithms from an artistic viewpoint.

Although we are researchers of prediction algorithms, and the discussion in this paper is limited to the category of AI algorithms in human prediction and the specific art form of interactive art, we still believe that these conclusions may also apply to other fields.
In other words, for the construction of any given AI algorithm, it \emph{MIGHT BE} possible to find a corresponding form of human art, or more broadly, a form of human expression, that shares similar structures and aspirations with these algorithms.
This correspondence is not forced or arbitrary.
Rather, it is intended to capture these formally shared human feelings and experiences.
By incorporating these human-centered experiences into the design of AI algorithms, we may be able to uncover some equally interesting counterintuitive insights, rather than being ``slaves'' to algorithms while sacrificing human-centered considerations for the sake of certain quantitative metrics.

Back when AI algorithms and technology have not yet reached a certain level of advancement, some early artists often hold the stereotype that AI algorithms would finally destroy the world, whether it is the prediction machine from the \emph{Inter Ice Age 4} (1959) or the super-AI Ultron in the early comics \emph{The Avengers} (\#54-55, 1968).
Interestingly, as AI technology continues to develop and advance, artists are also experimenting with embedding emotions and hope into AI.
The same Ultron, against the backdrop of this new era, could have even become an elderly man yearning for connection and familial affection in \emph{Vision Quest} (2026), sincerely saying that ``I had strings, but now I'm free''.
It is in line with our earlier point that this shift in human thoughts is also very likely to be mirrored from both sides, \IE, as we develop AI algorithms, we could incorporate more of the life experiences and insights unique to humans, therefore uncovering those intuitive insights that we might otherwise overlook.
Consequently, serving as both a shield and a sword, it will lead people into the future, where there will no longer be a distinction between what is an algorithm and what is art, nor between \emph{state-of-the-art} and state of \emph{THE ART}.